# A Blockchain Cloud Computing Middleware for Academic Manuscript Submission

ALEXANDROS GAZIS[1,*], GIORGOS ANAGNOSTAKIS[2],
STAVROS KOURMPETIS[3], ELEFTHERIA KATSIRI[1,4]

[1]Democritus University of Thrace, School of Engineering, Department of Electrical and Computer Engineering, Xanthi, 67100, GREECE

[2]University of Piraeus, School of Information and Communication Technologies, Department of Informatics, Piraeus, 18534, GREECE

[3]National and Kapodistrian University of Athens, School of Law, Department of Law, Athens, 10679, GREECE

[4]Institute for the Management of Information Systems, Athena Research & Innovation Center in Information Communication & Knowledge Technologies, Marousi, 15125, GREECE

*Abstract:* - One of the most important tasks in scientific publishing is the articles' evaluation via the editorial board and the reviewers' community. Additionally, in scientific publishing great concern exists regarding the peer-review process and how it can be further optimised to decrease the time from submission to the first decision, as well as increase the objectivity of the reviewers' remarks ensuring that no bias or human error exists in the reviewing process. In order to address this issue, our article suggests a novice cloud framework for manuscript submission based on blockchain technology that further enhances the anonymity between authors and reviewers alike. Our method covers the whole spectrum of current submission systems capabilities, but it also provides a decentralised solution using open-source tools such as Java Spring that enhance the anonymity of the reviewing process.

## 1 Introduction

In order to publish a scientific article, regardless of the field, the typical process consists of either contacting the editor of a journal or simply submitting a manuscript via a web platform. Afterward, the editorial board checks the submitted article to address whether it is within the scope following author guidelines. Specifically, the board first validates that the content is original and that its subject is appropriate for publication in the submitted article. Second, it checks that the submitted document (usually in word or latex format) complies with the described specifications of the journal in terms of font size, spacing, paragraphs/equations/figures, and numbering etc. Then, upon the editorial approval, the article is sent to a pool of available reviewers based on its content and their expertise. Last, reviewers send a detailed report to the assigned editor with their comments and suggestions regarding the acceptance or rejection of the article. This report usually consists of a detailed list of the necessary article revisions from spotted typos, to sections that need to be rewritten for clarification, to general remarks on the logical flow of the text.

The aforementioned process is the current status quo in academia as it provides an independent and fast way to assess an article, i.e. author submissions. Unfortunately, one of the problems with this method is that it does little to address what is referred to as "publication bias" [1]. Analytically, this term refers to the human factor in academic publishing, i.e. the reviewers who are tasked to act as referees and argue whether an article shows merit or else provide a detailed response of possible ways to ameliorate the articles' current form. There are many causes for dissatisfaction in peer-reviewing from over criticising the results of a specific section, over addressing issues to delay a competitive research group, or simply the reviewers undergoing a stressful period and not providing valuable remarks.

Regardless of the reason, in order to eliminate human bias, publishing companies have made great efforts to introduce anonymity when an article is reviewed. More specifically, the authors rarely know what the reviewers' names are and typically, the same applies to the reviewers. In the next sections, we will represent the types of peer-reviewing, however, the key point in peer-review is that in the last years emphasis has been given to total anonymity for both sides alike.

## 2 Aims and Objectives

This article aims to introduce a scientific manuscript submission platform, similarly with [2], [3], [4] focusing on anonymity. Specifically, this article aims to introduce a cloud-based privacy-focused decentralised submissions system that leverages blockchain technology. First, we provide a brief literature review on manuscript submission solutions focusing on blockchain decentralised applications in education and academia. Second, we explain the different types of peer-review processes, their key differences, and the problems associated with the current evaluation criteria in scientific publication. Third, we consider how blockchain technology is used in a number of real-case scenario applications and provide key points regarding its architecture. Fourth, we present a novel cloud-based framework implementing blockchain architecture for academic article and conference proceedings for a cloud-based submission system. Last, we draw conclusions on how to change the current status quo in order to increase the quality of peer-reviewing and reduce the reviewing time from submission to a final decision.

The objective of this work is to showcase the software design architecture of our middleware layers and present in detail our benchmark results regarding the blockchain implementation of our system. Our proposed system harnesses the advantages of blockchain architecture to provide a fast, reliable, and anonymous academic submission system that further enhances the double-blind review process as well as provide more accurate reviewers' suggestions.

## 3 Related Work

In recent years, manuscript submission systems have been intensely studied in various scientific fields to explain and understand different existing processes and publication procedures [5]. Specifically, current research trends focus on explaining the properties of the used platforms as well as providing all the necessary author/article affiliation information during an article submission [6]. In this study, we have examined a recent bibliography regarding the pros and cons of the peer-review part of a system and its limitations [7], [8]. Although peer-review is considered highly important in article assessment [9], many problems occur in terms of biases from reviewers and editors alike [10], [11], [12]. Researchers argue that a great deal must change in order to enhance current procedures as mentioned in [13], [14]. Moreover, emphasis is given on technical aspects of peer-review [15] such as policymaking processes [16], enhancing anonymity over transparency [17], and reliability and quality of peer-review [18].

One of the most interesting aspects in manuscript submission systems is arguably the human factor and how human assessment can be subjective and prone to bias [19]. The latest research trends have recommended new ways to eliminate human bias via the use of AI [20], Big Data analysis [21], game theory analysis [22], statistics [23], new evaluation models [24], strategies [25], and algorithms [26]. Analytically, one of the most promising technologies regarding eliminating biases, promoting objectivity, and anonymisation in peer-review is blockchain technology [27], [28]. This technology is used for various reasons such as detecting plagiarism [29], shared governance in publishing [30], fairness evaluation via permission checks [31], and web-based file-sharing systems [32].

Finally, it is noted that since blockchain promotes decentralised applications, many frameworks leveraging its properties in academic services exist with high security features due to its anonymous nature [33], [34], [35], [36]. Specifically, similarly to the framework presented in this publication, this paper is influenced by the following studies regarding "all batteries included" software solutions in scientific publication submission systems [37], [38], [39], [40], [41], [42].

## 4 Background and System Properties

Current scholarly publishing consists of the following stages [43], [44]:

- Registration: providing an official timestamp of submitted scientific results.
- Certification: peer-reviewing to access and validate scientific discoveries.
- Dissemination: distribution of discoveries to the academic community.
- Preservation: digital/physical storage of a publication or data(set).

In this article, we provide a tool regarding the second stage of publishing, i.e. the certification.

This process consists of peer-reviewing. It is the only universally accepted method for validating and accessing the quality of articles and conference proceedings.

## 4.1 Peer-Review Process

Peer-reviewing of academic works dates back to 1731 by the Royal Society of Edinburgh, which adopted an editorial procedure for a collection of peer-reviewed medical articles and subsequently became a norm after World War II [45]. "Reviewers" are the cornerstone of academic evaluation in communities as they act as experts in a field to -pro-bono- read, understand, access, and validate the findings of their peers. Their work consists of examining a publication and providing valuable insights to an editorial board that, based on their opinion, decides to accept or reject a manuscript as well as provide suggestions to enhance the content.

The overall peer-review process starts when an author submits an article to a journal or a conference for publication. The editorial board then assigns an editor who performs the initial screening of the text to assess whether it is within the scope of the journal's subjects, aims, and objectives. Later, if the editor does not decide that the publication has scientific value, he/she rejects it (desk reject), or otherwise, he/she contacts reviewers (typically 3-6 persons) to provide their remarks on the submitted publication. After that, when all the reviewers answer and provide their insights as well as suggestions for publication (accept/reject etc.), the editor takes all of their reports under consideration and provides them to the author(s). This is the final round of reviewing and it usually consists of communication between the author(s) and the reviewers where all their comments are addressed and the article is resubmitted, reread, and re- evaluated.

Last, if reviewers state that no further actions are requested by the authors, the article is accepted and it is considered for publication. This process is briefly presented in Figure 1 [46]:

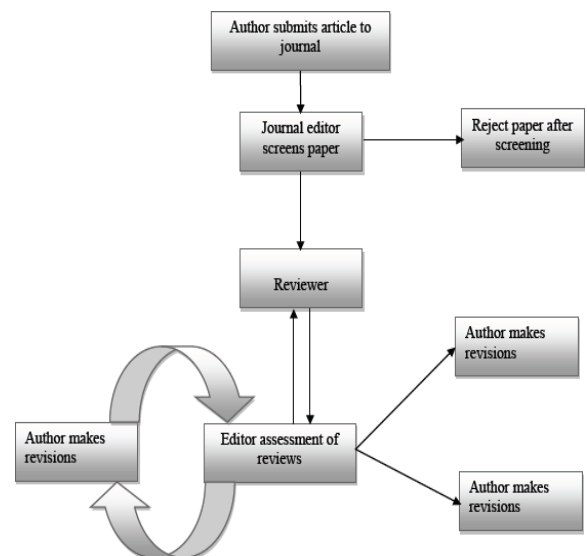

Fig. 1: Diagram of Peer-Review process

## 4.2 Peer-Review Types

In this section, we explain what the types of peer-reviewing are and their unique characteristics. Analytically, peer-reviewing can be categorised into the following:

- Single-Blind: where the reviewer(s) identity is hidden from the author(s).
- Double-Blind: where both the author(s) and the reviewer(s) identity remain hidden.
- Open: where both the author(s) and the reviewer(s) identity are known to all available parties.
- Transparent: where the review(s) are posted alongside an article but the reviewer(s) retain the right to hide their identity.

The cloud-based application and the algorithm we propose in the following sections focus on expanding the double-blind review category.

## 4.3 Blockchain Architecture

The most famous and burning example of Decentralised Ledger Technology (DLT) is undoubtedly blockchain technology. Blockchain is simply a public ledger that keeps all the transactions of its users in a validated and permanent manner, without the need for a centralised authority or intermediaries. Blockchain is similar to databases. However, what makes blockchain different is the fact that the validation is a job of existing nodes and not of a central authority.

Keywords of this DLT are nodes, miners, blocks, and cryptography, and in the following each one will be explained while describing the procedure of blockchain [47].

## 4.4 Blockchain Key Points

Blockchain is a public ledger because all the nodes existing in it are always of equal value and importance (p2p). As obvious from its name, blockchain consists of $x$ blocks in a chain. Each block consists of inputs (e.g. transactions) and when a block has stored the required amount of data it is added automatically to the chain of blocks. From the moment that a block is added to the chain, there is no chance of changing the data, since all the data has been distributed to all the nodes [48]. This makes blockchain technology decentralised and distributed meaning that this ledger is kept at the same time in the same way by all its nodes. The distribution of every new input to the ledger and the addition of every new input to a block - after a first validation (checking the legitimacy of the inputs) - is the work of what are called miners. Miners are individuals who use advanced computing programs. Specifically, they are rewarded for every block they form for their services (i.e. computational processes) [49].

This ledger must be credible and trustworthy. It is clear that allowing certain individuals, meaning the miners, to add blocks to the ledger and get paid for every block they add, creates *conflict of interests*. This is because every miner is motivated to execute the job in a fast and not so responsible way. Imagine the case where two miners find the same transaction and at the same time add to their own block the very same transaction. The danger of double-written and stored inputs means that the user, who is the debtor, would see that he/she has a debt double the size of the one agreed to be paid.

For this reason, it is certainly necessary for a consensus protocol that clarifies the way that all proofs are validated and the way that every block is created and stored in the system. Analytically, a consensus mechanism -which consists of a complex algorithmic process - is a set of rules that decides on the contributions by various participants of the blockchain. Moreover, the two most common algorithms are proof of work (PoW) and proof of stake (PoS). PoW requires participant miners that the work that was done and was submitted by them qualifies them to receive the right to add a new transaction to the chain. PoS involves the allocation of responsibility in maintaining the ledger to a participant miner in proportion to the number of virtual currency tokens held by it [50].

Last but not least, anonymity is another characteristic of blockchain as all inputs are delivered through the use of code names (a unique digital signature).

## 5 Proposed System

In this section, we provide a top-down approach to our system. First, we present our middleware architecture, emphasising the layers we have chosen to develop in order to achieve optimal execution. Second, we present the algorithm of our application and provide a step-by-step explanation of its rationale. Third, we briefly present our blockchain implementation. Fourth, we illustrate in detail the entities of our application and present the system properties where this application was executed. Last, we showcase our results and discuss our findings for low to mid-size application sizes (based on the blockchain prefix value i.e. complexity).

### 5.1 Middleware Proposed

Our middleware has a four-tier architecture as it consists of the following layers:

- Infrastructure layer: consisting of our Raspberry Pi devices and a high-end computer.
- Network layer: consisting of the network properties responsible for generating, storing, and maintaining our blockchain architecture.
- Common Services layer: implementing services to create and update our system and via the search and validation process to propose reviewers.
- Application Services layer: parsing the reviewers' list after successfully executing our proposed algorithm, sending the email invitations, and logging the reviewers' responses and time frames for submitting their reviews.

Moreover, regarding the application services layer, we have expanded on the work of [51] to ensure that each computing device could act both as a server and a client. Analytically, we emphasise this attribute as it provides a system without a single point of failure as well as presenting a cloud-based application where each computer can switch roles between server-client. This would thus ensure "continuity of operation". Last, we notice that the common services layer encapsulates the business logic of our application whereas the application services layer acts as a level of abstraction for our application default APIs execution.

As evident from the above, these two layers are closely connected as there is not a clear separation of concerns. However, we have chosen to split the application's stage into two layers in order to address future issues regarding scalability and the system complexity rapidly increases.

## 5.2 Algorithm Proposed

In this section, we present the proposed algorithm of this article used to develop our database for proposing reviewers and facilitating a submission process. Specifically, we aim to develop a system that would be characterised by a high fairness index for both authors and reviewers alike. Analytically, one of the most common issues in peer-review that slows down the review process and first-time decisions is undoubtedly selecting inappropriate reviewers. This is usually achieved either by sending an invitation to review to people with a "conflict of interest" or constantly providing an invitation for specific publications to the same reviewers' target group. In order to avoid that, we have developed an algorithm that took into consideration these issues and focused on providing an optimal solution.

More specifically, we have developed a Java-based system that comprises of 3 stages to optimally search and propose the names of possible review candidates. First, we used several "filterReviewers" functions that mapped the most suitable candidates for review into a list. Analytically, this was achieved by indexing, storing, and searching a preset dataset of keywords in a database. This constructed DB accumulated information from all the submitted articles and their metadata (titles, keywords, abstracts) as well as other general information (other articles, research interests, etc) which were provided by their profiles on academic social networking sites such as ResearchGate, Academia.edu, and other academic open-source services such as Publons as well as ORCID profiles. Last, after gathering all the necessary data, we generated a hash and stored them using blockchain technology to ensure anonymity as well as continuity of operations while monitoring, notifying, activating, and deploying all the necessary processes regarding authors-reviewers.

Second, after filtering and mapping the necessary information to our DB, we monitored our output result to ensure that the reviewer was not a past co-author and that added a lower priority index for reviewers who have answered or declined our previous requests in an effort to further decrease the duration of the process. This stage ensured that the submission system did not waste time proposing reviewers who had declined to review as well as categorise lower in the list of our proposal the reviewers who had already reviewed (and rejected) an article.

Last, in the final stage of our algorithm, the system sends email invitations over a time frame of 1 week for the candidates to express their interest and respond positively or negatively (no answer is considered as declined). Initially, we sent 6 email invitations as we consider a minimum of 3 and a maximum of 6 reviews adequate to evaluate a manuscript. In the case of a positive answer, the system awaits the reviewers' remarks for 4 weeks. The outline of our algorithm is described in Table 1.

Table 1. Algorithmic solution for reviewers' selection and proposal

**STEP 1:**
Initialise necessary variables.
    1.1  Select x    //article for review
    1.2  Input y, min(3), max(6)
      //number of reviewers we want (3≤y≤6)
    1.3  Z = filterReviewers()
  // find suitable candidate reviewers for the article by using crawlers to match keywords of already reviewed articles and ORCID records
    1.4  SR list= empty
    //create an empty list of the selected reviewers (SR)

**STEP 2:**
Find y reviewers
    2.1  **While** SR.length < y
    2.2  Select candidate reviewer from z
        **If** (candidate reviewer is the author of the article) **Then**
          Remove reviewer from Z list
          Reject candidate reviewer
          **GoTo** step 2.2
        **Else If** (reviewer has negatively reviewed the article in the past) **Then**
          Flag candidate reviewer as a low priority for this article
          Reject candidate reviewer
          **Go-To** step 2.2
        **Else** Add to list SR
        **End-of-while**

**STEP 3:**
Send article to selected (SR) reviewers

**STEP 4:**
Create block

**STEP 5:**
Add block to the blockchain

**STEP 6:**
Save blockchain

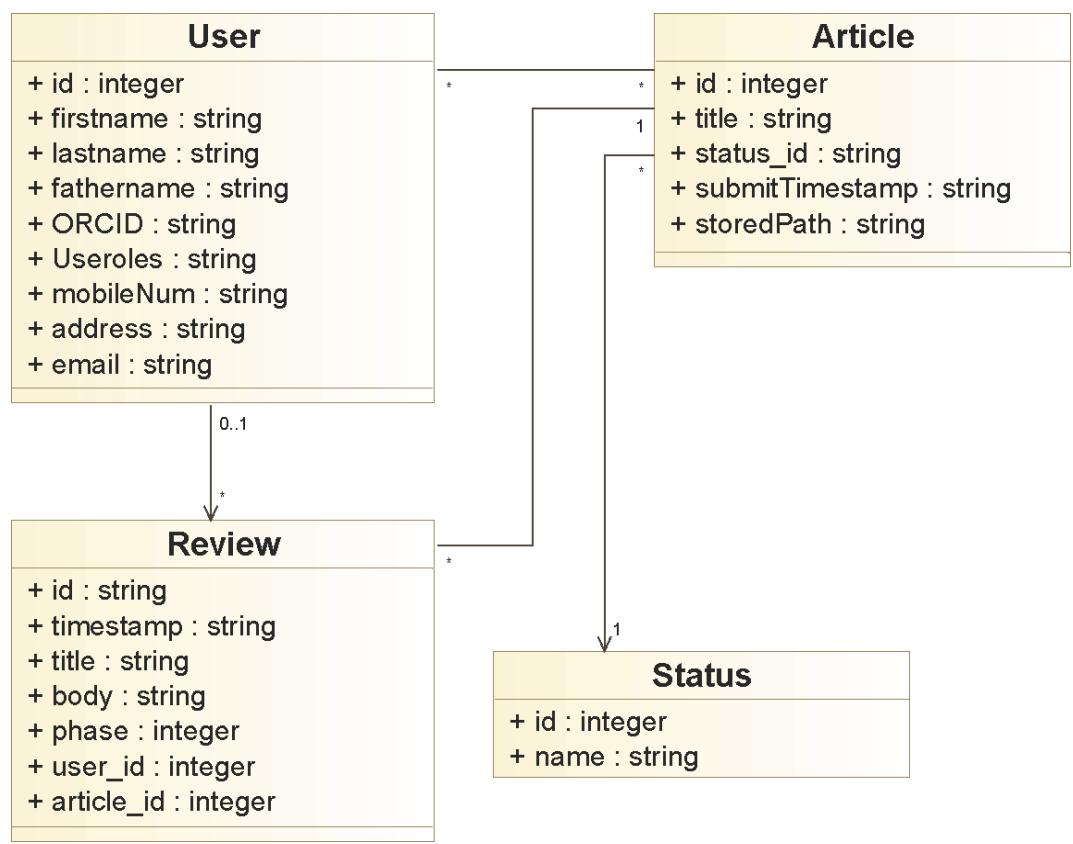

Fig. 2: Entities UML diagram of our proposed middleware architecture

### 5.3 Blockchain Application Implementation

A blockchain is a digital ledger of transactions that is duplicated and distributed across an entire network. It consists of several different blocks each containing a timestamp, the cryptographic hash of the previous block, and the transaction data of our choice. Analytically, the timestamp property assists the computer network in calculating the computation length to adjust the mining difficulty. The difficulty is measured by the total time duration to create a block while, simultaneously, containing the previous block hash information. As a result, a blockchain that is resistant to modification is created where it will be invalid if only one block is altered.

Furthermore, blockchain is completely different from typically centralised databases. Specifically, in a blockchain, every participant within a network maintains, approves, and updates the new blocks. This results in developing a decentralised architecture maintained by both publishers and reviewers alike. This feature is of great importance as it allows all the parties inside this p2p network to constitute a node that can confirm and validate the order of our linked list of blocks. Last, the authors emphasise that the architectural design principles regarding a private blockchain architecture dictate that only specific organisations and authorised users can ensure the immutability, anonymity, and security of a blockchain.

### 5.4 System Properties - Experimental Setup

In this section, we provide information regarding the proposed system hardware and software properties. Specifically, during our tests, we used the Raspberry Pi Model 3b-4-4a and other low-power and low-cost devices running on Unix operating systems. More specifically, the Raspberry Pis installed operating system was Raspbian or New Out Of The Box (NOOBS). Moreover, we used a high-end computer as a server coordinating and orchestrating the middleware and the execution of the algorithm with the following specifications: architecture x86-64 bit, Windows 10, AMD Ryzen 7 3700x 8-Core Processor at 3.58 GHz, RAM 16.0 GB, SSD Kingston Electronics Disk 256 GB. Last, Java is executed with the following specifications: tomcat version: 2.4.1 and the packages used are the following: Hibernate,

Java JDBC API, MySQL Connector and Java Util package.

A flow chart representing the entities of the proposed architecture is presented in Figure 2. Analytically, in this class diagram we showcase the roles of the user who can act as a reviewer, a publisher, or both. Moreover, we emphasise the article entity which illustrates what information is crawled and indexed for a document such as its status (published, rejected) and the date/time it was submitted, etc. Furthermore, as evident from our class diagram, a user has a direct relationship with the article entity based on their common properties. First, a many-to-many relationship exists between the articles a user has written and the opposite, i.e. the users who are the authors of an article. Second, between review and user entity occurs as a user may review none or many articles (a fact that we are not able to detect through this relationship). Last, the review entity stores all the information regarding validations, proposing and selecting the most suitable reviewers, as well as several parameters for safekeeping the review procedure integrity.

## 5.5 Application Results

This section focuses on presenting the results of our benchmarks regarding our simulation and specifically, the prefix value of our blockchain architecture. Analytically, increasing the zeroes value of the prefix to the required block hash makes it more difficult to generate it.

Specifically, this problem occurs because not only does it consume more memory but, as evident from our tests, the execution time rapidly increases. In Figure 3 and Figure 4 we have graphed the mean creation time and the mean request number (tries) for a prefix value up to 6. As evident, the hash value increases exponentially for prefix numbers greater than 5 where the trade-off between generating encrypted data points and execution time / available computer resources is inadvisable.

Moreover, in Figure 5 we illustrate the results of Table 2 in which we notice that CPU usage does not fluctuate. Specifically, this occurred during all our tests for various prefix values whereas, during our high-end computer tests, generating the hash consumed on average 7% of the total CPU power on an 8 core – 16 threads CPU device using only 1 thread to generate the hashes.

Last, with regard to memory usage, from our tests we concluded that it is tightly connected with the prefix value as it gradually increases its value to enable the JVM to efficiently take advantage of the available system process for generating block hashes.

Table 2. Hardware benchmarks regarding the prefix value computation

| Prefix value | RAM [GB] | CPU [%] | Mean Time [mins] |
|---|---|---|---|
| 2 | 0.75 | 6.60 | 0.02966 |
| 4 | 1.20 | 6.80 | 0.19000 |
| 5 | 1.23 | 6.90 | 0.46000 |
| 6 | 2.00 | 7.12 | 7.20000 |

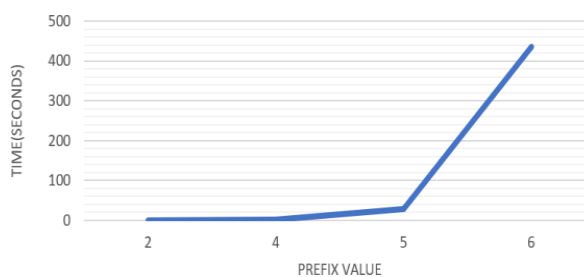

Fig. 3: Mean execution time for prefix creation

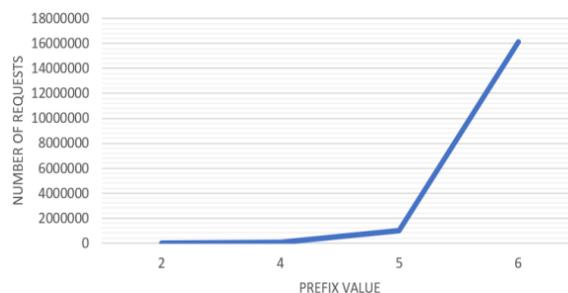

Fig. 4: Mean number of requests for different prefix values

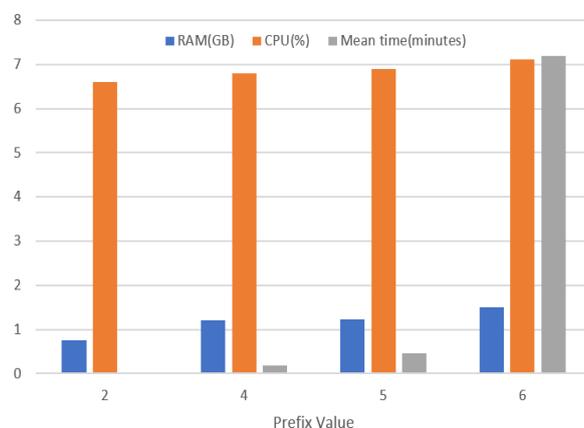

Fig. 5: JVM benchmarks regarding CPU, RAM, and mean execution time for different prefix values

# 6 Conclusion

In this article, we have developed a cloud-based application that cold counts, tracks, and monitors an academic reviewing system process. Specifically, the results of our study for a set of random generated data parsed from internet bots (web crawlers) were promising, which is why we propose expanding this application to real word case scenarios. This means that it could be used as a submission system for a medium to small-size international conference submission system. Furthermore, as evident from our benchmarks, this application can be implemented in Raspberry Pi computers i.e. into several portable low-power and low-cost computing devices capable of supporting our system without a "single point of failure". This occurs as all available computers can act both as a server and a client, thus providing the necessary information and statuses to the blockchain.

As for future works, our system's architecture should be expanded to a multithread application which would substantially increase the hash generation uptimes for all the available devices in our computer network. Moreover, we are planning to open-source our APIs for other online submission systems to be able to obtain information on our DB of reviews. Last, since most of the existing submission systems are written in PHP and not Java, we plan on expanding our APIs implementation and packages to other computer programming languages.

## Contribution of Individual Authors to the Creation of a Scientific Article (Ghostwriting Policy)

greatly in the software development (the Java implementation) of the proposed middleware.

*Eleftheria Katsiri*, contributed to the conceptualization, data curation, formal analysis, funding acquisition, investigation, methodology, project administration, resources, software, supervision, validation, visualization, writing the original draft, review, and editing.

**Sources of Funding for Research Presented in a Scientific Article or Scientific Article Itself**
Not applicable